# Electron transport in folded bilayer-bilayer graphene/hexagonal boron nitride superlattices under high magnetic fields

Takuya Iwasaki[1*], Motoi Kimata[2], Yoshifumi Morita[3*], Shu Nakaharai[1*], Yutaka Wakayama[1], Eiichiro Watanabe[4], Daiju Tsuya[4], Kenji Watanabe[5], Takashi Taniguchi[1], and Satoshi Moriyama[6*]

[1]International Center for Materials Nanoarchitectonics, National Institute for Materials Science (NIMS), 1-1 Namiki, Tsukuba, Ibaraki 305-0044, Japan

[2]Institute for Materials Research, Tohoku University, 2-1-1 Katahira, Aoba-ku, Sendai, Miyagi 980-8577, Japan

[3]Faculty of Engineering, Gunma University, Kiryu, Gunma 376-8515, Japan

[4]Nanofabrication Platform, NIMS, 1-2-1 Sengen, Tsukuba, Ibaraki 305-0047, Japan

[5]Research Center for Functional Materials, NIMS, 1-1 Namiki, Tsukuba, Ibaraki 305-0044, Japan

[6]Department of Electrical and Electronic Engineering, Tokyo Denki University, 5 Senju-Asahi-cho, Adachi-ku, Tokyo 120-8551, Japan

E-mail: IWASAKI.Takuya@nims.go.jp, morita@gunma-u.ac.jp, NAKAHARAI.Shu@nims.go.jp, moriyama.satoshi@mail.dendai.ac.jp

**Abstract**

Employing graphene as a template, we fabricate moiré superlattices by stacking bilayer or folded bilayer-bilayer graphene (BLG or fBBLG) and hexagonal boron nitride (hBN), i.e., hBN/BLG/hBN or hBN/fBBLG/hBN stacks, with a small twist angle between the graphene and one of the two hBN layers. Because of the modulation due to the hBN, higher-generation Dirac points can emerge with a narrow bandwidth and van Hove singularities. In the moiré superlattice devices, we can therefore access the higher-generation Dirac points by in-situ gate tuning. This study is based on our previous paper (Appl. Phys. Express **13**, 035003 (2020)). Here we show more extended data by applying high magnetic fields up to ~24 T. We also comment on the temperature dependence of the resistivity and magnetoresistance with reference to the 'plain' BLG data for a comparative study.

## 1. Introduction

Carbon-based superlattices form a novel class of quantum metamaterials, where carbon material for the template ranges from 1D to 2D, i.e., carbon nanotube[1] to graphene (Gr).[2,3] For example, graphene superlattices are fabricated by direct stacking of ultrathin/atomic-layer materials in contrast to 'classic' molecular beam epitaxy/pulsed laser deposition techniques.[4] For small twist angles in the stacking, the lattice mismatch can lead to a long-range spatial modulation —i.e., a moiré superlattice— which is now a basic technique to engineer energy band structures.[5,6] In graphene, higher-energy scales, e.g., in the vicinity of the van Hove singularity, are hard to access. With the advent of moiré superlattices, however, the energy band engineering, e.g., by using hexagonal boron nitride (hBN),[7–9] offers an unprecedented route by in-situ gate tuning. Here, we employ hBN/Gr/hBN (sometimes called Gr/hBN for simplicity) moiré superlattices by stacking bilayer or folded bilayer-bilayer graphene (BLG or fBBLG) and hBN, i.e., BLG/hBN or fBBLG/hBN stacks, with a small twist angle between the graphene and one of the two hBN layers. Because of the modulation due to the hBN, higher-generation Dirac points (or their descendants, e.g., in the case of BLG) can emerge with a narrow bandwidth. We can therefore access the higher-generation Dirac points by in-situ gate tuning. Here we note that fine-tuning to the magic angle with a vanishing velocity (flat band) is not necessarily crucial, as demonstrated below.

This is an update of the data in Ref. 10) where a basic recipe for the method of fabricating the fBBLG/hBN superlattices was presented. In this paper, we show more extended data, applying high magnetic fields up to ~24 T. Based on the data in Ref. 11), we also comment on resistivity-temperature ($R$-$T$) characteristics and magnetoresistance with reference to 'plain' BLG data for a comparative study.

## 2. Methods

Figure 1 shows our device structures. To fabricate the devices, BLG, fBBLG, and hBN flakes were firstly set on a $SiO_2$/Si substrate by the mechanical exfoliation method from bulk crystals. By the bubble-free dry transfer technique,[12] we assembled the hBN/fBBLG/hBN heterostructure, which was etched into a Hall bar geometry by electron beam (EB) lithography and $CHF_3/O_2$ plasma. Electrodes with one-dimensional edge contact (Cr/Pd/Au) were fabricated by the EB deposition.[9] Another hBN sheet was transferred onto the above device to protect the edge of the Hall bar. The top gate (Ti/Au) was then fabricated by EB lithography and deposition. The 'plain' BLG and BLG/hBN superlattice devices were also fabricated in the same manner. Note that the BLG/hBN superlattice was fabricated from the same stack with the fBBLG/hBN superlattice. The schematic cross-sectional view of the 'plain' BLG, 'plain' BLG/hBN superlattice, fBBLG/hBN superlattice are shown in Figs. 1(a)-1(c), respectively.

For the transport measurement, a four-terminal configuration with AC lock-in techniques (an excitation current $I \sim 10$–$100$ nA and a frequency of 17 Hz) was applied to detect the longitudinal resistivity $\rho_{xx} = (V_{xx}/I)(W/L)$ and the Hall resistivity $\rho_{xy} = V_{xy}/I$, where $L$ and $W$ are the channel length and width, respectively, $V_{xx}$ and $V_{xy}$ are the voltage measured between the Hall bar electrodes shown in Fig. 1(d). The device was set at low temperature ($T$) in a $^4$He cryostat with a superconducting magnet used for applying a perpendicular magnetic field ($B$). For the plain BLG device, the heavily doped Si substrate was used as the back gate. For the BLG/hBN superlattice device, a graphite flake under the bottom hBN layer was used as the back gate. For the fBBLG/hBN superlattice device, the top and back gates were used to modulate a vertical displacement field and the carrier density ($n$) of the device independently.

## 3. Basic results: gate-voltage dependence, *R-T* characteristics and magnetoresistance

After comments on 'plain' BLG (without superlattice structure/'satellites') (Fig. 1(a)) and 'plain' BLG/hBN superlattice (Fig. 1(b)) as comparative studies, let us move on to our focus in this paper, i.e., fBBLG/hBN superlattice (Fig. 1(c)).

### 3.1 'Plain' BLG (without superlattice structure/'satellites')[8]

Data of the 'plain' BLG (Fig. 1(a)) are shown for a comparative study. Please note that, although the hBN/BLG/hBN structure is employed in this device, we do not observe 'satellites' within our window and we basically identify this with a BLG device of high quality. Figure 2(a) shows $\rho_{xx}$ and $\rho_{xy}$ as a function of the back-gate voltage ($V_{bg}$) at $T = 1.6$ K. The $\rho_{xx}$ peak and the $\rho_{xy}$ sign change at $V_{bg} \sim 0.2$ V corresponds to the signal of the charge neutrality point (CNP), where the carrier type is switched between electrons and holes. Figure 2(b) shows the mapping plot of $\rho_{xx}$ as a function of $V_{bg}$ and $B$ at $T = 1.6$ K in a log scale. As a result of Landau quantization, the constant-filling-factor regions dispersing from the CNP with increasing $B$ are well-resolved. The $T$-dependence of $\rho_{xx}$ is shown in Figs. 2(c)-2(e). In graphene, it has been well-established that the temperature dependence of the resistance is basically due to phonons.[13,14] The magnetoresistance is shown in Figs. 2(f) and 2(g). In the weak $B$ limit, as discussed in Ref. 15–17), Anderson localization effects due to impurity scatterings can play a role generally in graphene. Here geometric effects due to boundary scattering can be overlapped. A crossover takes place toward the quantum Hall effect in the strong $B$ limit. The plateaus are observed at $\rho_{xy} = h/(4Ne^2)$ ($N$, $h$, and $e$ represent an integer, a Planck constant, and an elementary charge, respectively), which imply the Landau quantization sequence of BLG.[18]

### 3.2 'Plain' BLG/hBN Superlattice[19–22]

The next focus is on the BLG/hBN superlattice device (Fig. 1(b)). In the very beginning, let us confirm its moiré effects. Figure 3(a) shows $\rho_{xx}$ and $\rho_{xy}$ of the BLG/hBN superlattice device as a function of $V_{bg}$ at $T = 1.7$ K. The $\rho_{xx}$ peak and the $\rho_{xy}$ sign change at $V_{bg} \sim -0.2$ V corresponds to the CNP of this device, while those at $V_{bg} \sim -15.7$ V and $V_{bg} \sim 15.3$ V reflect the hole- and electron-side 'satellites' respectively, which is induced by the moiré superlattice structure and leads to the estimate of the moiré wavelength and the stacking angle to be ~10.7 nm and ~0.85°, respectively.[22,23] Figure 3(b) shows the $\rho_{xx}$ mapping plot as a function of $V_{bg}$ and $B$ at $T = 1.7$ K. This shows the Landau fan spectra from the CNP and the satellites up to $B = 24$ T. Concerning the magnetoresistance shown in Fig. 3(c), Anderson localization effects such as the decrease of $\rho_{xx}$ with increasing $B$ could be observed in the weak $B$ limit. An interplay of several characteristic times is crucial to fix the magnitude of such quantum interference effects.[15-17] For the strong $B$ region, the magnetoresistance is non-monotonically changed with decreasing $n$ (increasing the hole density) due to the overlap with the spectrum from the hole-side satellite, in contrast to that in the BLG device shown in Fig. 2(g). Figure 3(d) shows the $\rho_{xx}$ as a function of $n$ for various $T$, which is converted into the $R$-$T$ characteristics shown in Figs. 3(e)-3(h), where $n$ is estimated by the parallel plate capacitance model. Let us consider the scattering mechanisms for the "normal" phase. In graphene, it is generally established that the temperature dependence of the resistance is basically due to phonons,[13,14] which leads to the $T$-linear dependence of $\rho_{xx}$ as observed in Fig. 3(e) at, e.g., $n \sim -5 \times 10^{12}$ cm$^{-2}$. On the other hand, it has been reported that Coulomb interactions in graphene superlattices can dominate the scattering mechanism for the resistance through the Umklapp processes, which implies the $T^2$-dependence of $\rho_{xx}$[24,25] and should apply around $n \sim -2 \times 10^{12}$ cm$^{-2}$ in Fig. 3(g). For example, at $n = -2.2 \times 10^{12}$ cm$^{-2}$, a fitting of $\Delta\rho_{xx} = \rho_{xx}(T) - \rho_{xx}(T = 1.7$ K$)$ to a curve $\Delta\rho_{xx} \sim \alpha T^{\beta}$ ($\alpha$ and $\beta$ are fitting parameters) gives an exponent $\beta = 2.1$ (~2) and the exponent is kept to be $\beta \sim 2$ within the window $-2.2 < n < -1.1$ ($\times 10^{12}$ cm$^{-2}$).

### 3.3 fBBLG/hBN superlattice[10]

Here we discuss the transport properties in the fBBLG/hBN superlattice device (Fig. 1(c)). Figure 4(a) shows $\rho_{xx}$ and $\rho_{xy}$ of this device as a function of the top-gate voltage ($V_{tg}$) at $T = 1.7$ K. The $\rho_{xx}$ peak and the $\rho_{xy}$ sign change at $V_{tg} \sim 0$ V correspond to the CNP ($n \sim 0$) of this device. Figure 4(b) shows $\rho_{xx}$ as a function of $n$ for various $T$. The $\rho_{xx}$ peaks at $V_{tg} \sim -8.9$ V ($n \sim -4.2 \times 10^{12}$ cm$^{-2}$) and $V_{tg} \sim 7.9$ V ($n \sim 3.7 \times 10^{12}$ cm$^{-2}$) are assigned to the BLG/hBN moiré effect as discussed below.

Figures 4(c)-4(f) show the $R$-$T$ characteristics of this device for the various $n$ ranges. Let us comment on the scattering mechanisms for the "normal" phase, where the temperature dependence of the resistance is at least partly due to phonons.[13,14] Furthermore, different types of bosonic modes, e.g., critical fluctuations might play another role here,[10] combined with the Coulomb interaction/Umklapp process discussed in the above subsection. Another key ingredient for the scattering mechanism is a relative angle between two hBN layers in hBN/Gr/hBN stacks, which also controls the electron-phonon coupling. These should be closely related to the resistance 'bump' and 'drop' discussed in Ref. 10), which are sensitive to the device quality and a relative angle between graphene and hBN (and two hBN layers). Figures 4(g)-4(j) show the magnetoresistance in the weak $B$ region. Unconventional 'V-shaped' large magnetoresistance is also observed in this device (Figs. 4(h) and 4(i) at $T = 1.7$ K). As shown in Fig. 4, it can be fragile under 'perturbation' like a gate voltage (Figs. 4(g) and 4(j)) and a temperature (Fig. 4(i) at $T$ = 10 K). Careful assignment of the scattering mechanism and the origin of the magnetoresistance is left as a future study.

Transport properties of the fBBLG/hBN superlattice device under high magnetic fields are also shown in Fig. 5. Again, we confirm that the $\rho_{xx}$ peak and the $\rho_{xy}$ sign change at $V_{tg}$ ~ 0 V correspond to the CNP of this device. The $\rho_{xx}$ peak positions at $V_{tg}$ ~ –9 V and ~ 8 V could be translated into $n$ ~ –4.2 × $10^{12}$ cm$^{-2}$ and ~ 3.7 × $10^{12}$ cm$^{-2}$, respectively, which are close to the estimate by the data of the BLG/hBN superlattice made from the same hBN/Gr/hBN stack. Therefore, these peaks were assigned to the BLG/hBN moiré effect.[10] The BLG/hBN effect-induced hole-side $\rho_{xx}$ peak is basically fixed at $V_{tg}$ ~ –9 V even with the higher magnetic fields (~17 T). As shown in Fig. 5, a complex pattern, which is reminiscent of the Hofstadter butterfly,[26] is observed as a function of $B$.

## 4. Summary

We employed Gr/hBN moiré superlattices by stacking graphene and hBN (hBN/Gr/hBN stacks) with a small twist angle between the graphene and one of the two hBN layers, and we accessed the higher-generation Dirac points by in-situ gate tuning. This paper is an update of the data in Ref. 10) where a basic recipe for the fabrication was presented. We showed more extended data by applying high magnetic fields up to ~24 T. In addition, we commented on the $R$-$T$ characteristics and magnetoresistance. Fine hybrid quantum device structures, e.g., quantum point contact and single-electron transistor on the 'monolithic' Gr/hBN superlattices, are left for a future study.[27–32] Search for signatures of 'hidden' orders and genuine quantum-limited physics should also be pursued beyond strongly correlated/'Mott' physics.

**Acknowledgements**

The authors thank H. Osato from the NIMS Nanofabrication platform for assisting with the device fabrication. This work was partially supported by GIMRT Program with proposal No. 202012-HMKPA-0050 of the Institute for Materials Research, Tohoku University, Japan Society for the Promotion of Science (JSPS) KAKENHI Grant Number JP21H01400, A3 Foresight by the JSPS, NIMS Nanofabrication Platform in Nanotechnology Platform Project, and the World Premier International Research Center Initiative on Materials Nanoarchitectonics sponsored by the Ministry of Education, Culture, Sports, Science and Technology (MEXT), Japan.

**References**

1) S. Iijima, Nature **354**, 56 (1991).
2) K. S. Novoselov, A. K. Geim, S. V. Morozov, D. Jiang, Y. Zhang, S. V. Dubonos, I. V. Grigorieva, and A. A. Firsov, Science **306**, 666 (2004).
3) K. S. Novoselov, A. K. Geim, S. V. Morozov, D. Jiang, M. I. Katsnelson, I. V. Grigorieva, S. V. Dubonos, and A. A. Firsov, Nature **438**, 197 (2005).
4) L. Esaki and R. Tsu, IBM J. Res. Dev. **14**, 61 (1970).
5) R. Bistritzer and A. H. MacDonald, Proc. Natl. Acad. Sci. U.S.A. **108**, 12233 (2011).
6) R. Bistritzer and A. H. MacDonald, Phys. Rev. B **84**, 035440 (2011).
7) K. Watanabe, T. Taniguchi, and H. Kanda, Nat. Mater. **3**, 404 (2004).
8) C. R. Dean, A. F. Young, I. Meric, C. Lee, L. Wang, S. Sorgenfrei, K. Watanabe, T. Taniguchi, P. Kim, K. L. Shepard, and J. Hone, Nat. Nanotechnol. **5**, 722 (2010).
9) L. Wang, I. Meric, P. Y. Huang, Q. Gao, Y. Gao, H. Tran, T. Taniguchi, K. Watanabe, L. M. Campos, D. A. Muller, J. Guo, P. Kim, J. Hone, K. L. Shepard, and C. R. Dean, Science **342**, 614 (2013).
10) T. Iwasaki, Y. Morita, S. Nakaharai, Y. Wakayama, E. Watanabe, D. Tsuya, K. Watanabe, T. Taniguchi, and S. Moriyama, Appl. Phys. Express **13**, 035003 (2020).
11) T. Iwasaki, M. Kimata, Y. Morita, S. Nakaharai, Y. Wakayama, E. Watanabe, D. Tsuya, K. Watanabe, T. Taniguchi, and S. Moriyama, Ext. Abstr. Solid State Devices and Materials, 2021, p. 445.
12) T. Iwasaki, K. Endo, E. Watanabe, D. Tsuya, Y. Morita, S. Nakaharai, Y. Noguchi, Y. Wakayama, K. Watanabe, T. Taniguchi, and S. Moriyama, ACS Appl. Mater. Interfaces **12**, 8533 (2020).
13) E. H. Hwang and S. Das Sarma, Phys. Rev. B **77**, 115449 (2008).
14) S. Das Sarma, S. Adam, E. H. Hwang, and E. Rossi, Rev. Mod. Phys. **83**, 407 (2011).
15) F. V. Tikhonenko, A. A. Kozikov, A. K. Savchenko, and R. V. Gorbachev, Phys. Rev. Lett. **103**, 226801 (2009).
16) T. Iwasaki, S Nakamura, O. G. Agbonlahor, M. Muruganathan, M. Akabori, Y. Morita, S. Moriyama, S. Ogawa, Y. Wakayama, H. Mizuta, and S. Nakaharai, Carbon **175**, 87 (2021).
17) T. Iwasaki, S. Moriyama, N. F. Ahmad, K. Komatsu, K. Watanabe, T. Taniguchi, Y. Wakayama, A. M. Hashim, Y. Morita, and S. Nakaharai, Sci. Rep. **11**, 18845 (2021).
18) K. S. Novoselov, E. McCann, S. V. Morozov, V. I. Fal'ko, M. I. Katsnelson, U. Zeitler, D. Jiang, F. Schedin, and A. K. Geim, Nat. Phys. **2**, 177 (2006).
19) L. A. Ponomarenko, R. V. Gorbachev, G. L. Yu, D. C. Elias, R. Jalil, A. A. Patel, A. Mishchenko, A. S. Mayorov, C. R. Woods, J. R. Wallbank, M. Mucha-Kruczynski, B. A. Piot, M. Potemski, I. V. Grigorieva, K. S. Novoselov, F. Guinea, V. I. Fal'ko, and A. K. Geim, Nature **497**, 594 (2013).
20) K. Komatsu, Y. Morita, E. Watanabe, D. Tsuya, K. Watanabe, T. Taniguchi, and S. Moriyama, Sci. Adv. **4**, eaaq0194 (2018).

21) K. Endo, K. Komatsu, T. Iwasaki, E. Watanabe, D. Tsuya, K. Watanabe, T. Taniguchi, Y. Noguchi, Y. Wakayama, Y. Morita, and S. Moriyama, Appl. Phys. Lett. **114**, 243105 (2019).

22) C. R. Dean, L. Wang, P. Maher, C. Forsythe, F. Ghahari, Y. Gao, J. Katoch, M. Ishigami, P. Moon, M. Koshino, T. Taniguchi, K.Watanabe, K. L. Shepard, J. Hone, and P. Kim, Nature **497**, 598 (2013).

23) M. Yankowitz, J. Xue, D. Cormode, J. D. Sanchez-Yamagishi, K. Watanabe, T. Taniguchi, P. Jarillo-Herrero, P. Jacquod, and B. J. LeRoy, Nat. Phys. **8**, 382 (2012).

24) K. Kadowaki and S. B. Woods, Solid State Commun. **58**, 507 (1986).

25) J. R. Wallbank, R. Krishna Kumar, M. Holwill, Z. Wang, G. H. Auton, J. Birkbeck, A. Mishchenko, L. A. Ponomarenko, K. Watanabe, T. Taniguchi, K. S. Novoselov, I. L. Aleiner, A. K. Geim, and V. I. Fal'ko, Nat. Phys. **15**, 32 (2019).

26) D. R. Hofstadter, Phys. Rev. B **14**, 2239 (1976).

27) S. Nakaharai, J. R. Williams, and C. M. Marcus, Phys. Rev. Lett. **107**, 036602 (2011).

28) N. F. Ahmad, K. Komatsu, T. Iwasaki, K. Watanabe, T. Taniguchi, H. Mizuta, Y. Wakayama, A. M. Hashim, Y. Morita, S. Moriyama, and S. Nakaharai, Appl. Phys. Lett. **114**, 023101 (2019).

29) N. F. Ahmad, T. Iwasaki, K. Komatsu, K. Watanabe, T. Taniguchi, H. Mizuta, Y. Wakayama, A. M. Hashim, Y. Morita, S. Moriyama, and S. Nakaharai, Sci. Rep. **9**, 3031 (2019).

30) L. Veyrat, A. Jordan, K. Zimmermann, F. Gay, K. Watanabe, T. Taniguchi, H. Sellier, and B. Sacépé, Nano Lett. **19**, 635 (2019).

31) T. Iwasaki, S. Nakaharai, Y. Wakayama, K. Watanabe, T. Taniguchi, Y. Morita, and S. Moriyama, Nano Lett **20**, 2551 (2020).

32) T. Iwasaki, T. Kato, H. Ito, K. Watanabe, T. Taniguchi, Y. Wakayama, T. Hatano, and S. Moriyama, Jpn. J. Appl. Phys. **59**, 024001 (2020).

**Figure captions**

Figure 1. Schematic illustration of the device structures in this study. (a-c) Cross-section of the devices including (a) plain BLG, (b) plain BLG/hBN superlattice, and (c) fBBLG/hBN superlattice. For the superlattice devices (b) and (c), the crystallographic axis of BLG is aligned to that of one of the two (top and bottom) hBN layers with an angle of ~0 degrees. (d) Hall bar geometry and measurement configuration.

Figure 2. Transport properties in the plain BLG. (a) $\rho_{xx}$ ($B = 0$ T) and $\rho_{xy}$ ($B = 0.2$ T) as a function of $V_{bg}$ at $T = 1.6$ K. (b) $\rho_{xx}$ mapping as a function of $V_{bg}$ and $B$ at $T = 1.6$ K. (c) $\rho_{xx}$ as a function of $n$ for various $T$. (d,e) $\rho_{xx}$ as a function of $T$ for (d) various $n$, and (e) the CNP. (f) $\rho_{xx}$ and $\rho_{xy}$ as a function of $B$ at $T = 1.6$ K, $n = -1.0 \times 10^{12}$ cm$^{-2}$. (g) $\rho_{xx}$ as a function of $B$ at $T = 1.6$ K for various $n$.

Figure 3. Transport properties in the plain BLG/hBN superlattice. (a) $\rho_{xx}$ ($B = 0$ T) and $\rho_{xy}$ ($B = 0.2$ T) as a function of $V_{bg}$ at $T = 40$ mK. (b) $\rho_{xx}$ mapping as a function of $V_{bg}$ and $B$ at $T = 1.7$ K. (c) $\rho_{xx}$ as a function of $B$ at $T = 1.7$ K for various $n$. (d) $\rho_{xx}$ as a function of $n$ for various $T$. (e-h) $\rho_{xx}$ as a function of $T$ for various $n$.

Figure 4. Transport properties in the fBBLG/hBN superlattice. (a) $\rho_{xx}$ ($B = 0$ T) and $\rho_{xy}$ ($B = 0.22$ T) as a function of $V_{tg}$ at $T = 1.7$ K. (b) $\rho_{xx}$ as a function of $n$ for various $T$. (c-f) $\rho_{xx}$ as a function of $T$ for various $n$. (g-j) $\rho_{xx}$ as a function of $B$ at for various $T$ and $n$. $\rho_{xx}$ ($B$) for the negative $B$ is plotted as $\rho_{xx}$ ($-B$).

Figure 5. Magnetotransport in the fBBLG/hBN superlattice under high magnetic fields. (a) $\rho_{xx}$ and (b) $\rho_{xy}$ mapping as a function of $V_{tg}$ and $B$ at $T = 1.7$ K. The dotted lines are a guide for the eye, which is extrapolated from the spectrum dispersing from the CNP in the same measurement of the plain BLG/hBN superlattice device, shown in Fig. 3(b).

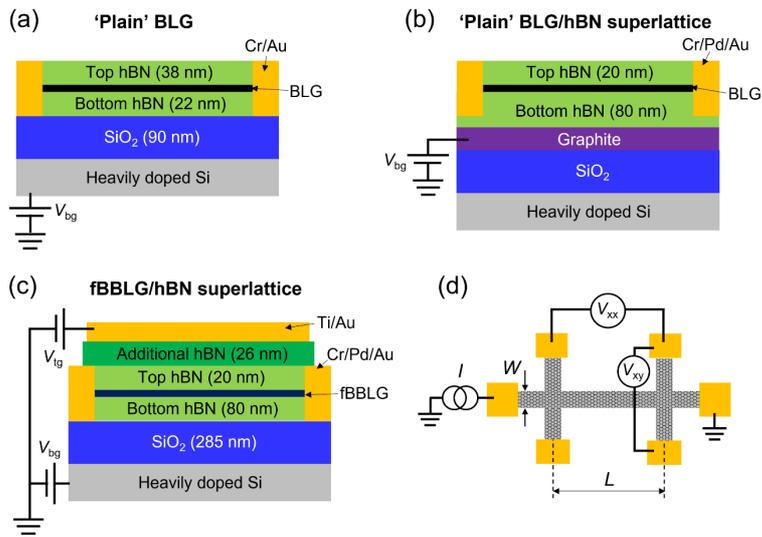

Fig. 1

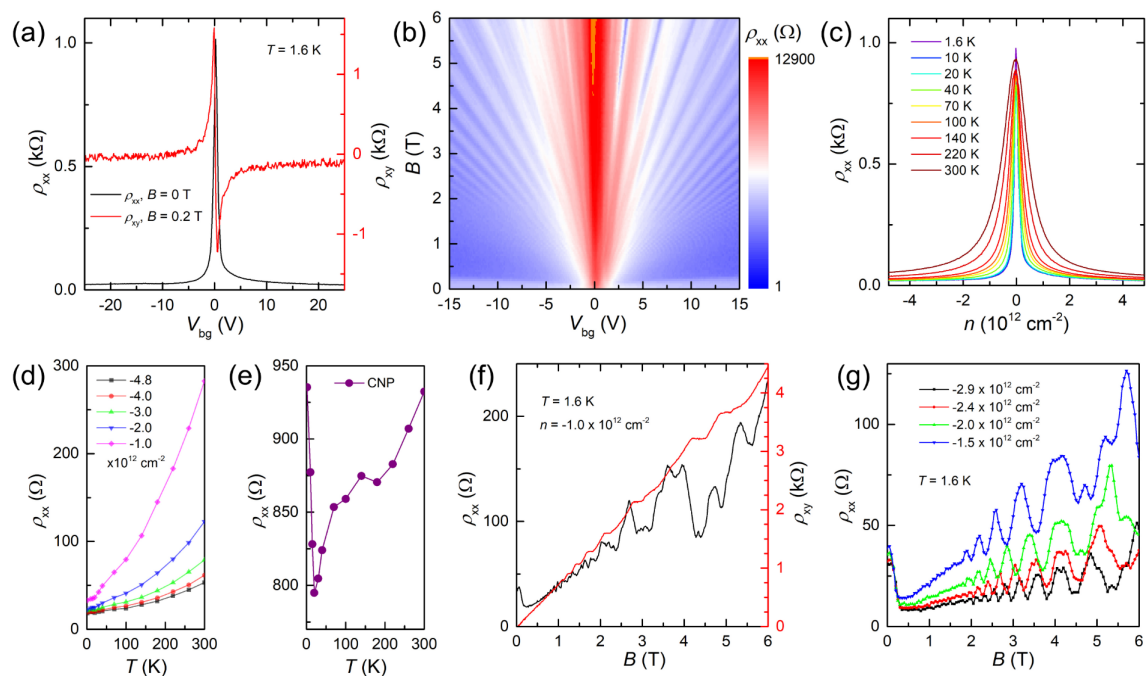

Fig. 2

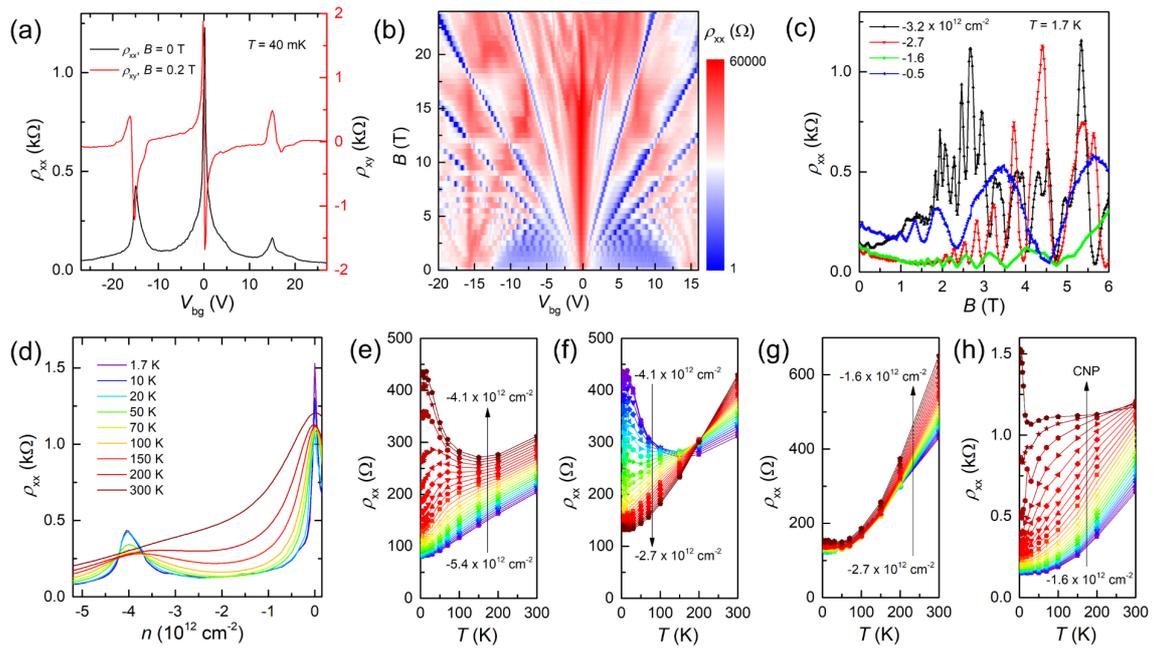

Fig. 3

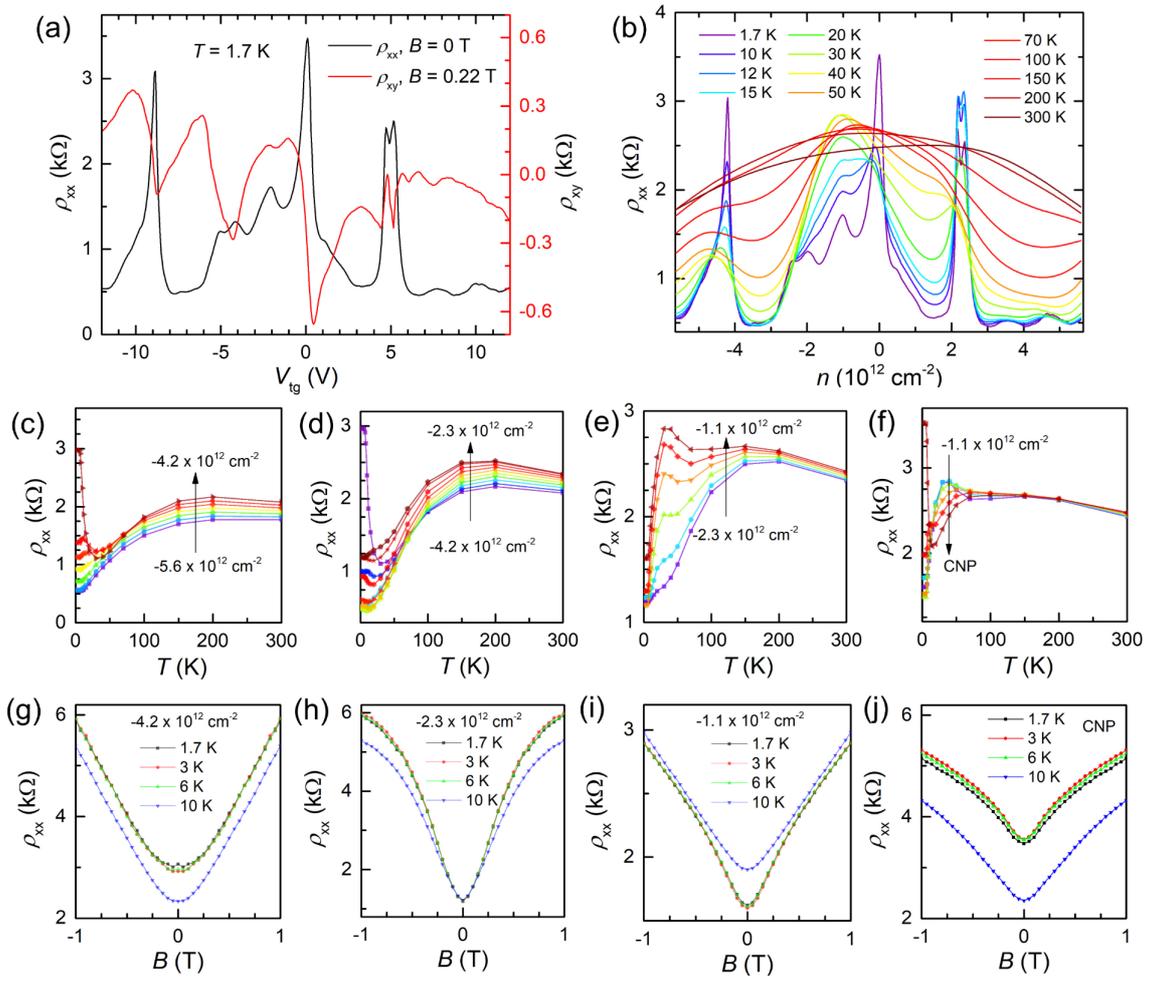

Fig. 4

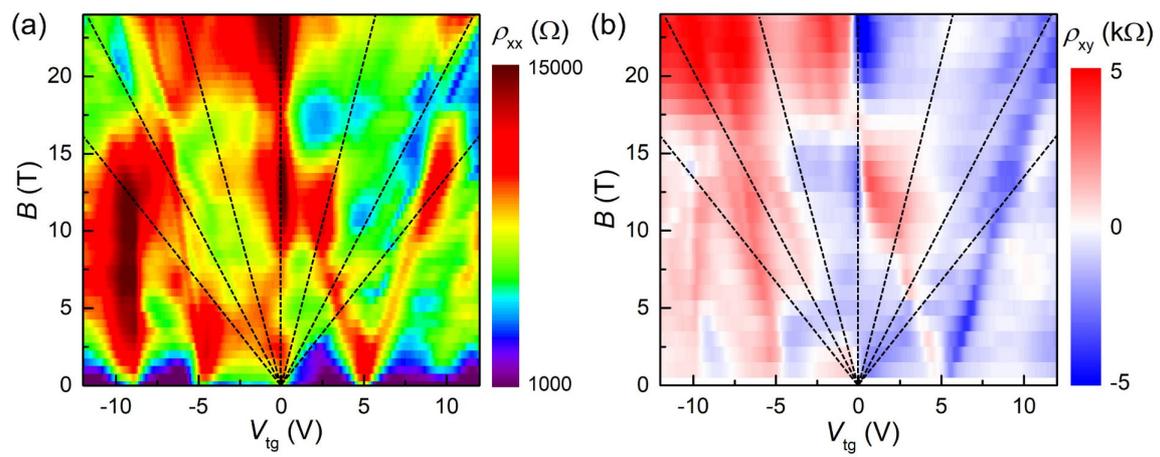

Fig. 5